\documentclass[12pt]{iopart}

\usepackage{graphicx}

\begin{document}

\title[]{Quarkonia Production in Cold and Hot Matters}

\author{Rapha\"el Granier de Cassagnac}

\address{Laboratoire Leprince-Ringuet, \'Ecole polytechnique/IN2P3, Palaiseau, 91128 France}
\ead{raphael@in2p3.fr}
\begin{abstract}
Quarkonia were predicted to be suppressed in the ``hot'' deconfined matter known as the Quark Gluon Plasma (QGP), but they were also seen to suffer from ``cold'' nuclear matter effects (parton shadowing, nuclear absorption...). Both at SPS and RHIC, suppression beyond nuclear effects was observed, but the rapidity dependence of the RHIC result is not easy to interpret. I review here the current status of these results, their possible interpretations and the new measurements that could provide insights on quarkonia suppression. Some of them were presented at this conference.
\end{abstract}


\section{A normal introduction}

Quarkonia suppression, as it was introduced by Matsui and Satz in 1986~\cite{Matsui:1986dk}, should be a direct signature of deconfinement.
Indeed, these heavy quark-antiquark bound states, the lighter of which being the $J/\psi$ particle, should melt in the quark gluon plasma thanks to the screening of the colour charge. Moreover, various bound states having different binding energies, they should melt at various temperatures. Measuring different quarkonia ($J/\psi$, $\psi'$, $\chi_c$, the $\Upsilon$ and $\chi_b$  families) could thus serve as a thermometer of the produced medium, each quarkonia suddenly disappearing above its proper dissociation temperature. However, it was soon realized that
if the hot quark gluon plasma could indeed melt quarkonia, cold normal nuclear matter can also absorbe them.

\begin{figure}
\begin{center}
  \begin{tabular}{cc}
  \includegraphics[width=6.8cm]{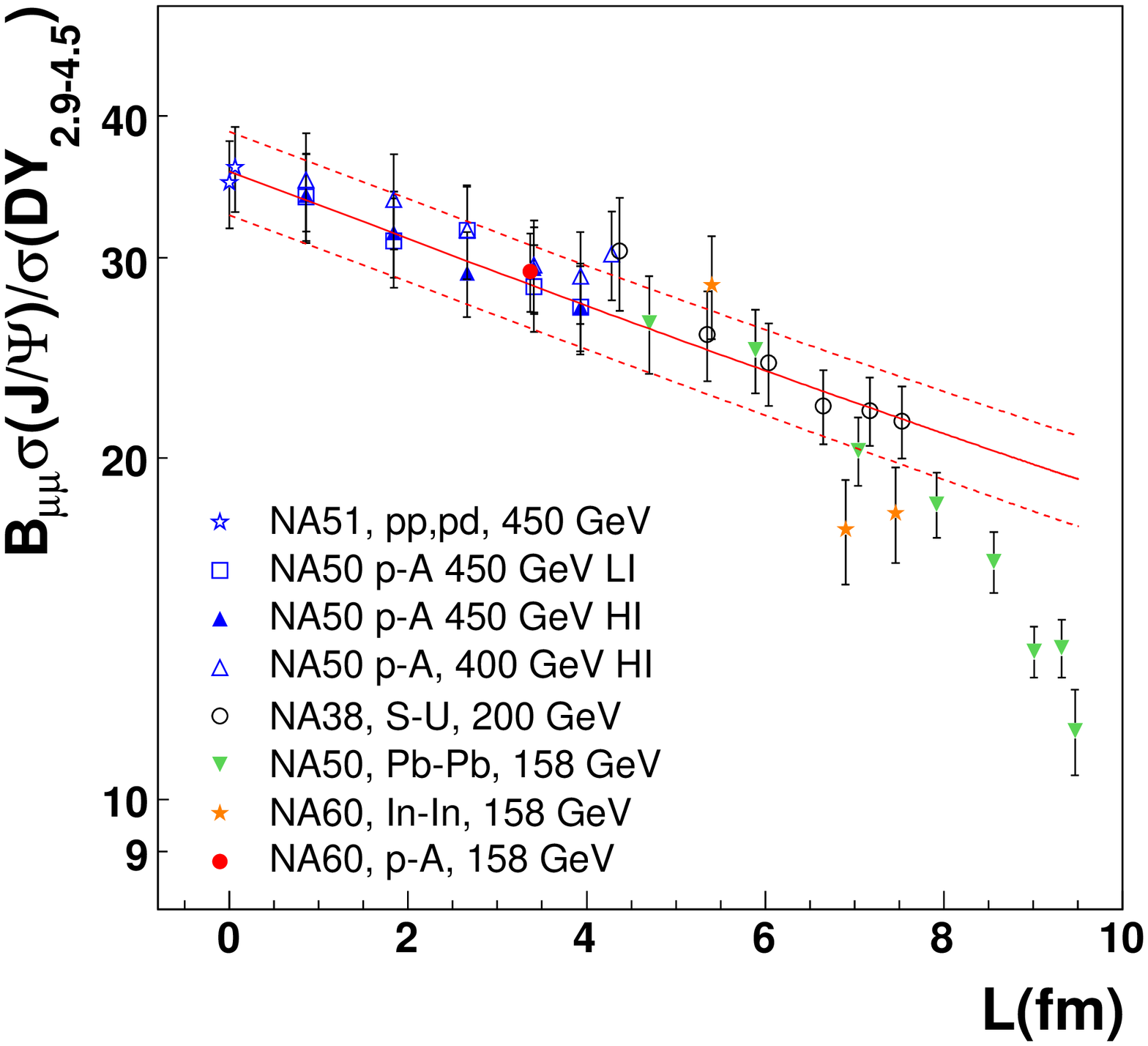} & \includegraphics[width=6.8cm]{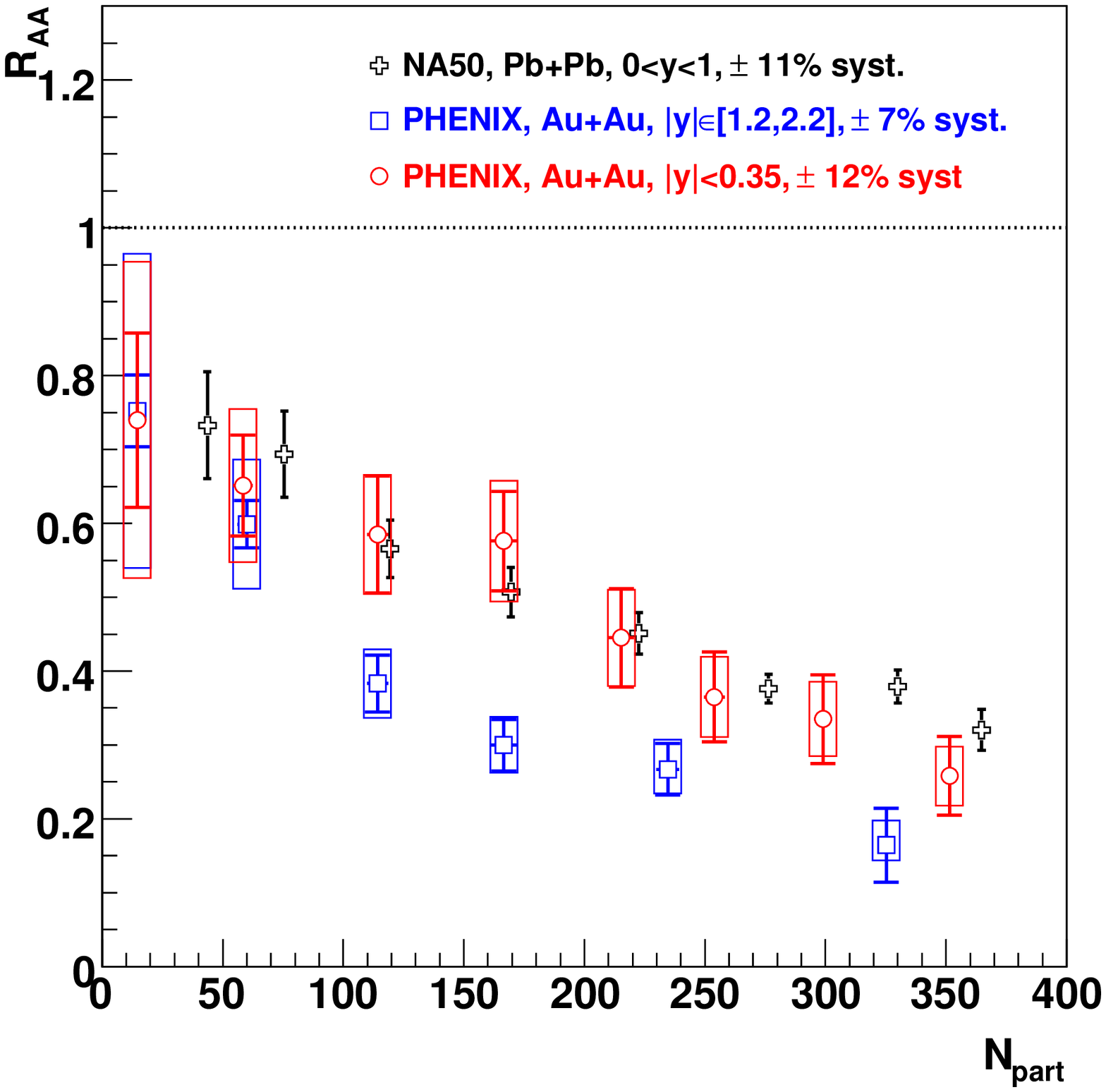}
  \end{tabular}
\end{center}
\vspace{-2ex} \caption{Left: $J/\psi$ yields normalized by Drell-Yan, as a function of the nuclear thickness $L$, as measured at the SPS~\cite{Scomparin:2007rt}. Right: $J/\psi$ nuclear modification factors of the hottest SPS (Pb-Pb) and RHIC (Au-Au) collisions, as a function of the number of participants $N_{part}$.
} \label{Fig1}
\vspace{-2ex}
\end{figure}

The left part of figure~\ref{Fig1} summarizes the $J/\psi$ yields (normalized by the Drell-Yan process which is known and observed to scale with binary collisions) measured at the CERN SPS. From p-p, to various p-A, to S-U, and up to peripheral In-In or Pb-Pb collisions, one can describe the data by a simple model in which the $J/\psi$ is absorbed by the forthcoming nucleons (their number being characterized through the nuclear thickness parameter $L$), with a corresponding cross-section of $4.18 \pm 0.35$~mb~\cite{Alessandro:2004ap}. Both the In-In and Pb-Pb more central collisions exhibit further \emph{anomalous} suppression.
Moreover, the precise measurements performed by the NA60 collaboration on In-In data show that the anomalous suppression saturates in the most central collisions~\cite{Arnaldi:2007zz}. Furthermore, the amount of anomalous suppression is compatible with the fraction of $J/\psi$ coming from excited states (30-40~\%, from $\psi'$ and $\chi_c$). One can thus think that, at SPS energies, namely around $\sqrt{s_{NN}}\simeq 20$~GeV, charmonia behave exactly like the predicted quark gluon plasma signature: a fast meltdown of some quarkonia, after a simple absorption cross section has been taken into account.

\section{The RHIC anomalies}

At RHIC energies, namely $\sqrt{s_{NN}}\simeq 200$~GeV, measurements of $J/\psi$ suppression in Au-Au collisions by the PHENIX experiment~\cite{Adare:2006ns} brought up two surprises, as it is shown on the right part of figure~\ref{Fig1}. First, at midrapidity (red circles), the amount of suppression is surprisingly similar to the one observed at SPS (black crosses) if plotted as a function of the number of participants $N_{part}$. There is no fundamental reason for this to happen, since the energy density should be higher at RHIC, and the cold nuclear effects could be drastically different (in particular, the initial gluons should be further shadowed, having smaller Bjorken $x$). Even more surprising is the fact that, at forward rapidity (blue squares), $J/\psi$ are further suppressed (by approximately 40\%). This is also confirmed by a preliminary analysis of the run~7 data shown at this conference.


So far, we think of two possible explanations of the RHIC $J/\psi$ data.

\begin{itemize}
\item First, $J/\psi$ could be indeed more suppressed than at SPS, but then recreated during (or at the term of) the hot partonic phase from initially uncorrelated $c$ and $\overline{c}$ quarks, the total number of initial $c \overline{c}$ pairs being larger than 10 in the most central collisions~\cite{Adler:2004ta,Abelev:2006db}. A large variety of such ``coalescence'' or ``recombination'' models exists, the latest flavours of which can be found in~\cite{Andronic:QM08,Thews:2005fs,Ravagli:2007xx,Zhao:2007hh,Yan:2006ve,Linnyk:2008uf,Tywoniuk:QM08}\footnote{For seminal publications see references therein. References~\cite{Andronic:QM08} and~\cite{Tywoniuk:QM08} were shown at this conference.}. Qualitatively, they explain why we see less suppression at midrapidity by the simple fact that there is more $c$ and $\overline{c}$ quarks to recombine there.
\item Second, $J/\psi$ could be more suppressed at forward rapidity because of nuclear effects. Standard gluon shadowing parametrizations do not tend to produce such a behaviour~\cite{Adare:2007gn} but they are poorly constrained by data and further saturation effects are not excluded. Interesting exploratory work to derive $J/\psi$ production in the framework of the Color Glass Condensate was shown at this conference~\cite{Nardi:QM08}. It is noticeable that, if such a cold matter effect is responsible for the different suppressions observed at mid and forward rapidity, it is possible that even at RHIC energy, only the excited charmonia states ($\psi'$ and $\chi_c$) are melting.
\end{itemize}

At this moment, one cannot exclude one or the other of the above scenarios. To draw a conclusion, more information is needed and the following section is a comprehensive list of observables, some of them having been shown at this conference, that will help moving forward in understanding $J/\psi$ anomalies seen at RHIC

\section{Moving forward}

\subsection{Open charm}

Currently, the open charm absolute cross-section is very poorly known at RHIC. This is unfortunate, since it is a basic input to recombination models, the probability to reproduce a quarkonia in the QGP depending quadratically on the number of initially produced $c \overline{c}$ pairs. The topic of open charm was covered in another contribution~\cite{Zhang:QM08} and will not be discussed further here. It is to be noted that the adjunction of new vertex detection capabilities in PHENIX and STAR will probably permit much better open charm and beauty measurements at RHIC in the future.

\subsection{Cold nuclear matter effects from d-Au}
\label{SecCNM}

\begin{figure}
\begin{center}
  \begin{tabular}{cc}
  \includegraphics[width=6.8cm]{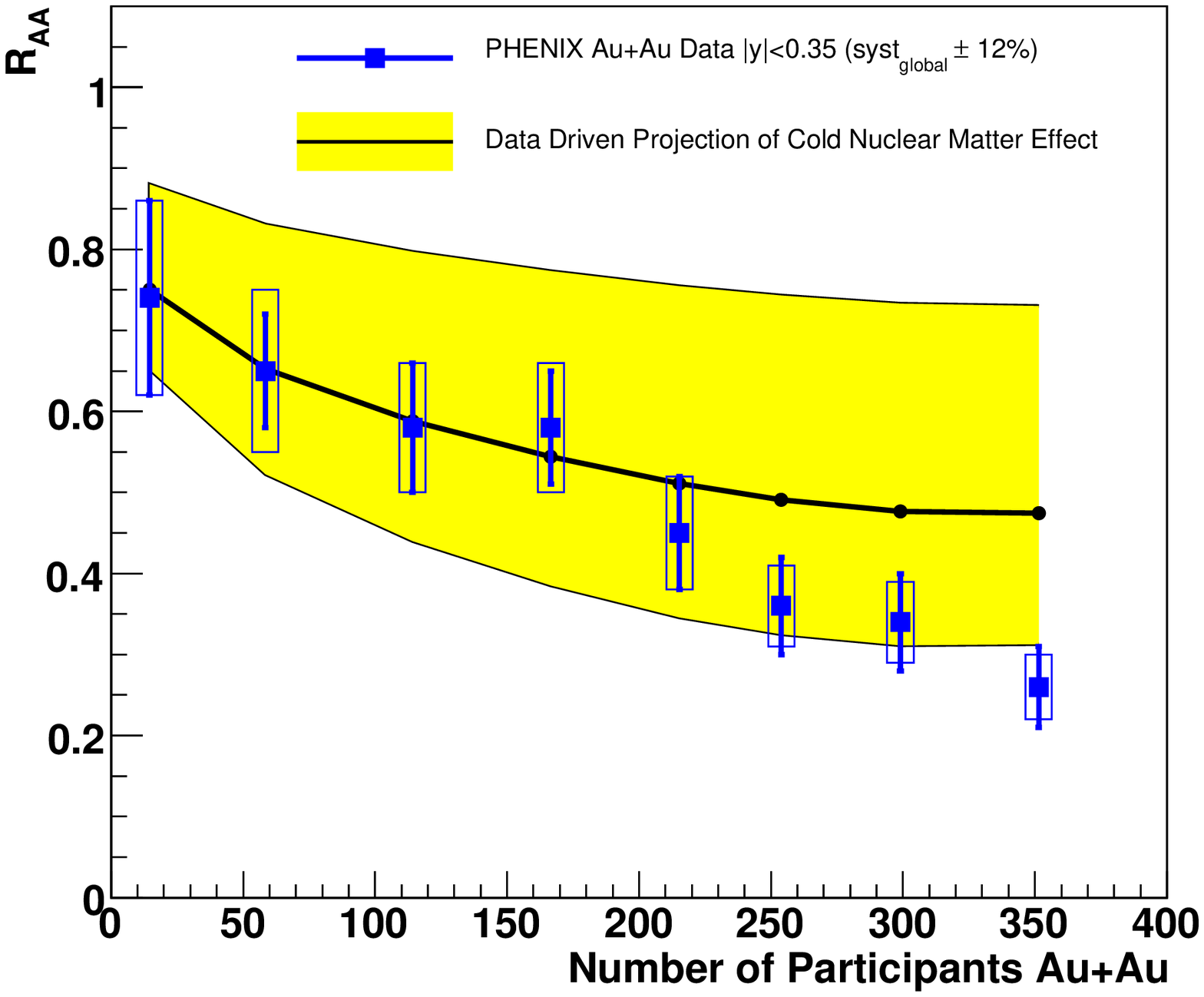} & \includegraphics[width=6.8cm]{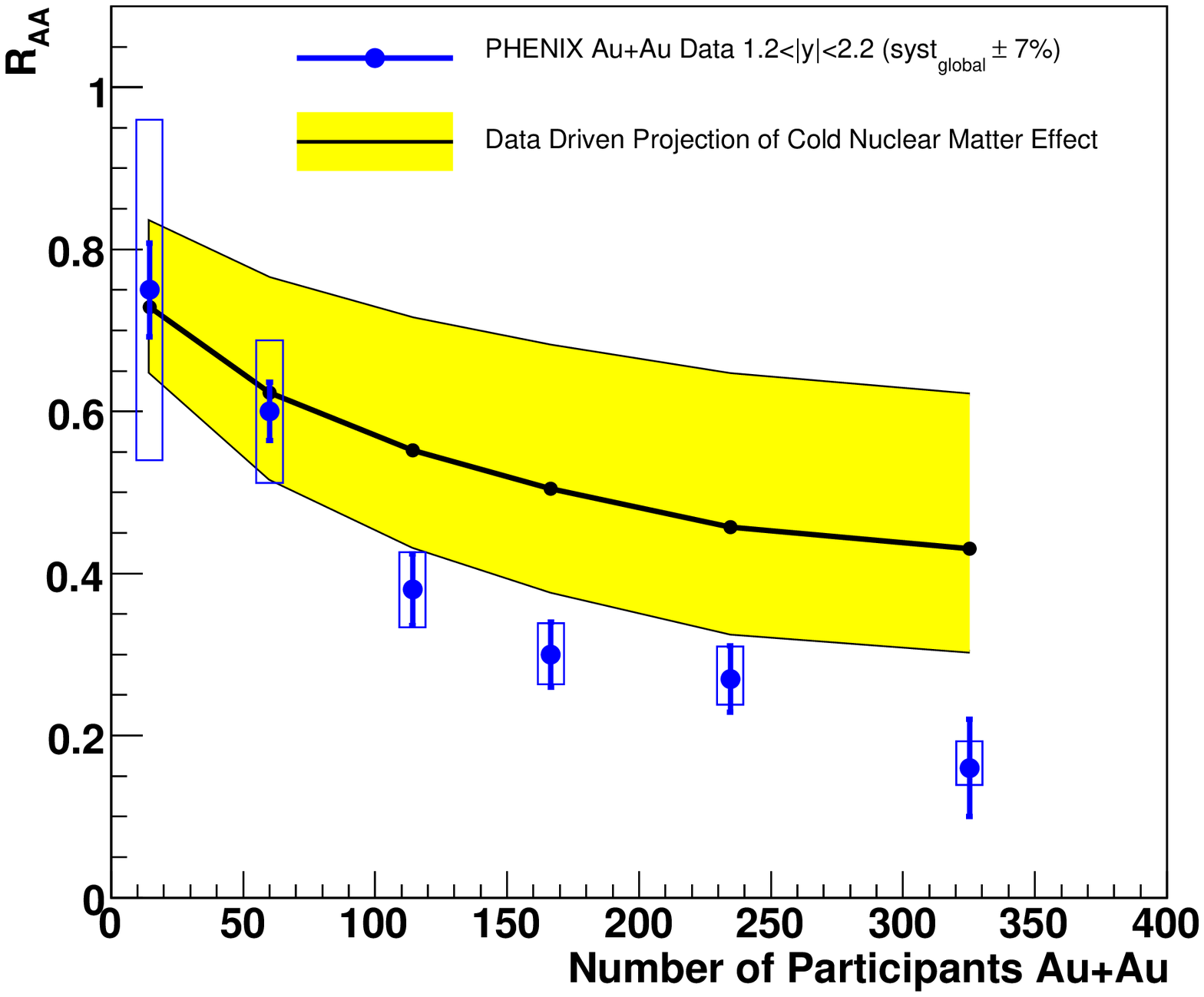}
  \end{tabular}
\end{center}
\vspace{-2ex} \caption{$J/\psi$ nuclear modification factors measured by PHENIX at mid (left) and forward (right) rapidity, compared to cold nuclear matter extrapolations from~\cite{Adare:2007gn}.
} \label{Fig2}
\vspace{-2ex}
\end{figure}

The effects of cold nuclear matter on $J/\psi$ production are also insufficiently known at RHIC. Extrapolations from the run~3 d-Au data set were discussed by the PHENIX collaboration in~\cite{Adare:2007gn} and at this conference~\cite{Wysocki:QM08}. Assuming given shadowing schemes, namely EKS98~\cite{Eskola:1998df} and NDSG~\cite{deFlorian:2003qf}, and propagating all the uncertainties from the \mbox{d-Au} measurements, one derives absorption cross sections of
$\sigma_{EKS} = 2.8^{+1.7}_{-1.4}$~mb or
$\sigma_{NDSG} = 2.2^{+1.6}_{-1.5}$~mb.
This illustrates that, at present, the uncertainty on such a break-up cross section is by no mean better than a couple of millibarns. Furthermore, in a data driven approach inspired by~\cite{GranierdeCassagnac:2007aj}, PHENIX shows that, if one prefers not to rely on a shadowing scheme and to instead propagate the centrality dependence measured in d-Au collisions to Au-Au (through a simple Glauber model), the amount of suppression due to cold nuclear matter is not know better than shown on figure~\ref{Fig2} by the shaded yellow bands. Such large uncertainties allow the anomalous suppression to be identical at mid and forward rapidity. For instance, the survival probabilities (beyond normal suppression) I derive from figure~\ref{Fig2} are $55^{+23}_{-38}$~\% and $38^{+18}_{-22}$~\% for the most central collisions, at mid and forward rapidity respectively. These numbers illustrate that there is an anomalous suppression, at least at forward rapidity. They also underline the crucial need for a more precise estimation of the cold nuclear effects. Hopefully, it will come from the large \mbox{d-Au} data sample recently produced by RHIC, the recorded luminosity being thirty times larger than the one the extrapolations of figure~\ref{Fig2} are based upon.

\subsection{Transverse momentum}

\begin{figure}
\begin{center}
  \begin{tabular}{cc}
  \includegraphics[height=6cm]{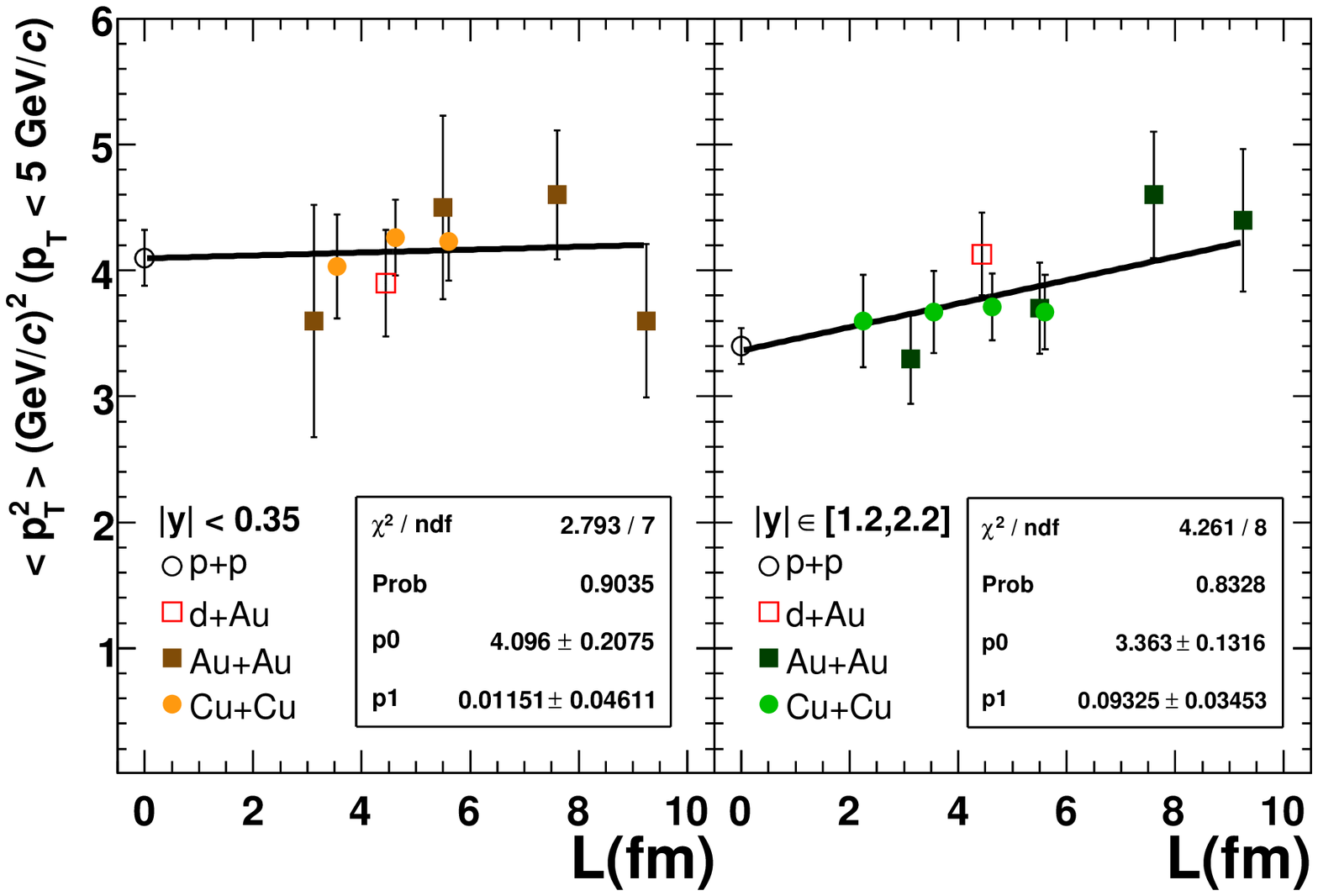}
  & \hspace{-3em} \includegraphics[height=6cm]{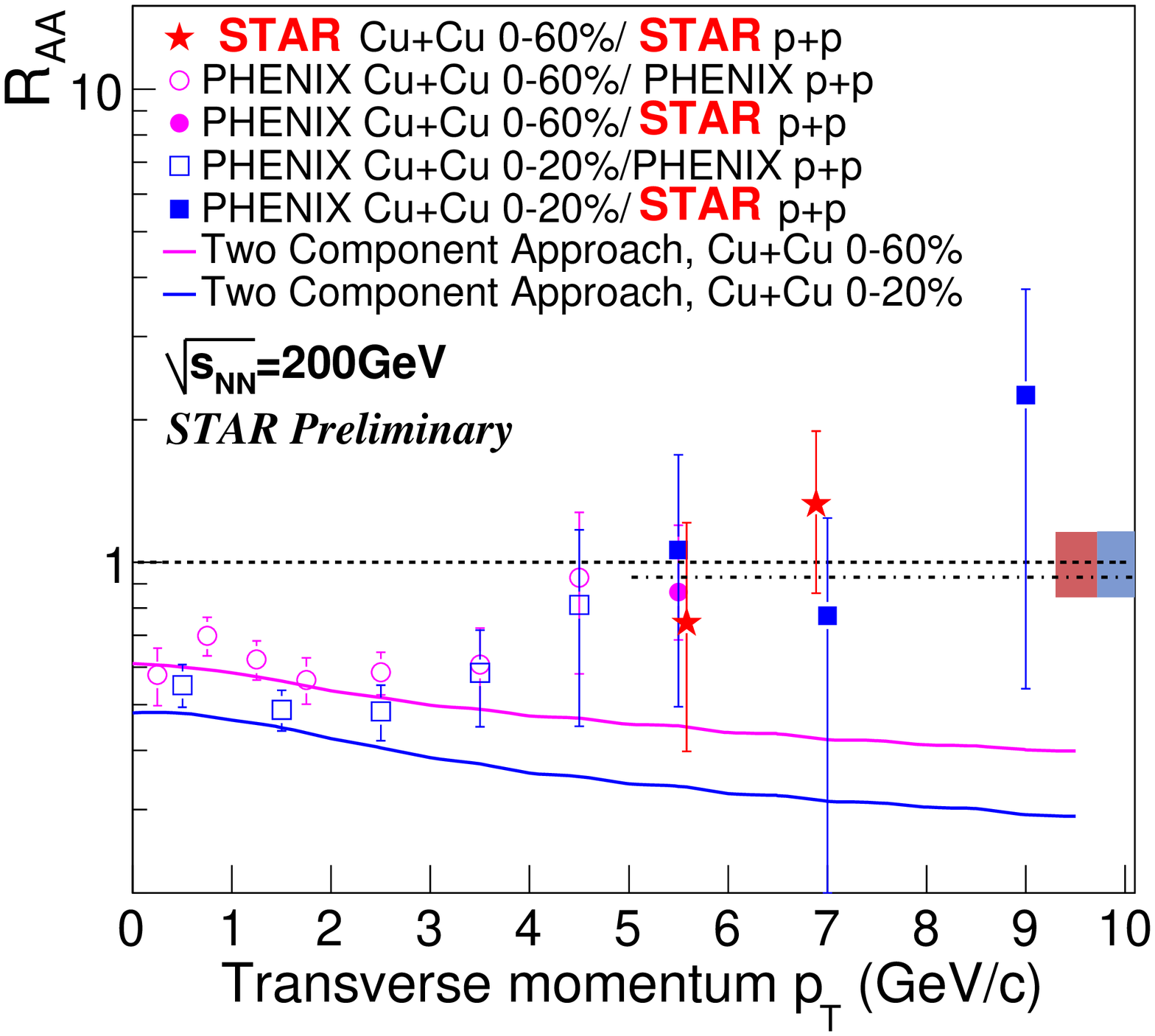}
  \end{tabular}
\end{center}
\vspace{-2ex} \caption{Left: $J/\psi$ average $<p_T^2>$ versus $L$ from PHENIX. Right: $J/\psi$ nuclear modification factor in Cu-Cu collisions from PHENIX (low $p_T$) and STAR (high $p_T$).
} \label{Fig3}
\vspace{-2ex}
\end{figure}

Since the yield of recombined quarkonia crucially depends on the (unknown) number of initially produced $c \overline{c}$ pairs, the magnitude of the nuclear modification factor itself is not an ideal probe of recombination. Instead, the evolution of other observables with respect to centrality could carry the signatures of recombined and/or initially produced $J/\psi$.
We already saw that narrowing the rapidity profile could be attributed to recombination.
Flattening the transverse momentum would be another of its manifestation. Indeed, the spectrum of $J/\psi$ formed from uncorrelated, randomly picked, $c$ and $\overline{c}$ quarks should be softer than the one of initially produced $J/\psi$. Here again, cold nuclear effects should play a significant role, by raising the initial parton $p_T$ through multiple scattering (Cronin effect), or even through shadowing \cite{Ferreiro:2008qj}. In case of pure Cronin effect, the average $p_T$ squared should vary proportionally to the amount of centers the partons can scatter upon, that can be characterized by $L$, the thickness parameter\footnote{Indeed, in the limit of instantaneous processes, the average number of subsequent or preceding interactions are proportional, and the same $L$ thickness parameter can be considered for initial (Cronin) of final (nuclear absorption) effects.}. This behavior was seen in various p-A collisions at various energies at SPS. During this meeting, the NA60 experiment~\cite{Arnaldi:QM08} reported a surprising preliminary result exhibiting a slope change between p-A at 158~AGeV and previous results, including A-A collisions at the same energy. This needs to be confirmed. The left part of figure~\ref{Fig3} shows the current status of similar measurements at RHIC. Simple linear fits to all the points, including p-p, d-Au, Cu-Cu and Au-Au data, exhibit excellent $\chi^2$ probabilities and, within these limited statistics, there is no evidence for a deviation from the Cronin effect, if any. The mid rapidity slope is even compatible with zero while the forward slope has a significance of 2.7~standard deviations. Their difference is not significant (1.4~standard deviations) and deserves more precise data, which will come soon for d-Au (run~8) and Au-Au (run~7).

In the same spirit, looking at higher $p_T$ may provide hints of the processes at play in $J/\psi$ production. On the left part of figure~\ref{Fig3}, PHENIX average $p_T^2$ were quoted for 0-5~GeV/$c$ intervals for consistency, but higher $p_T$ was reached in the most central collisions. At this conference, the STAR collaboration showed a $J/\psi$ signal (two standard deviations) in Cu-Cu collisions for $p_T$ above 5~GeV/$c$, as well as a more significant peak from \mbox{p-p} collisions~\cite{Tang:QM08}. While suffering from large uncertainties, the nuclear modification factor derived from these measurements is compatible with one, as shown by the red stars on the right part of figure~\ref{Fig3}, together with low $p_T$ corresponding PHENIX measurement (empty squares and circles). Statistically correlated high $p_T$ measurements derived from combining PHENIX and STAR are also displayed (full circles and squares). Before to draw any conclusions from this, one should remember that this behaviour was observed with high statistics at SPS\footnote{Hints can also be guessed in the PHENIX Au-Au~\cite{Adare:2006ns}, Cu-Cu~\cite{Adare:2008sh} and d-Au~\cite{Adare:2007gn} data, below 5~GeV/$c$.} for Pb-Pb~\cite{Ramello:2006db} and now for In-In collisions~\cite{Arnaldi:QM08}. In particular, one should keep in mind that the Cronin effect may justify this behaviour and will be soon investigated by both STAR and PHENIX with the high statistics d-Au data set now available at RHIC.

The most interesting result of these high $p_T$ $J/\psi$ from STAR, is a clear away side correlation with moderate $p_T$ hadrons ($>500$~MeV/$c$) seen in p-p collisions~\cite{Tang:QM08} that should start shading some light on the $J/\psi$ production mechanisms.


\subsection{Elliptic flow}

\begin{figure}
\begin{center}
  \begin{tabular}{cc}
  \includegraphics[height=4.6cm]{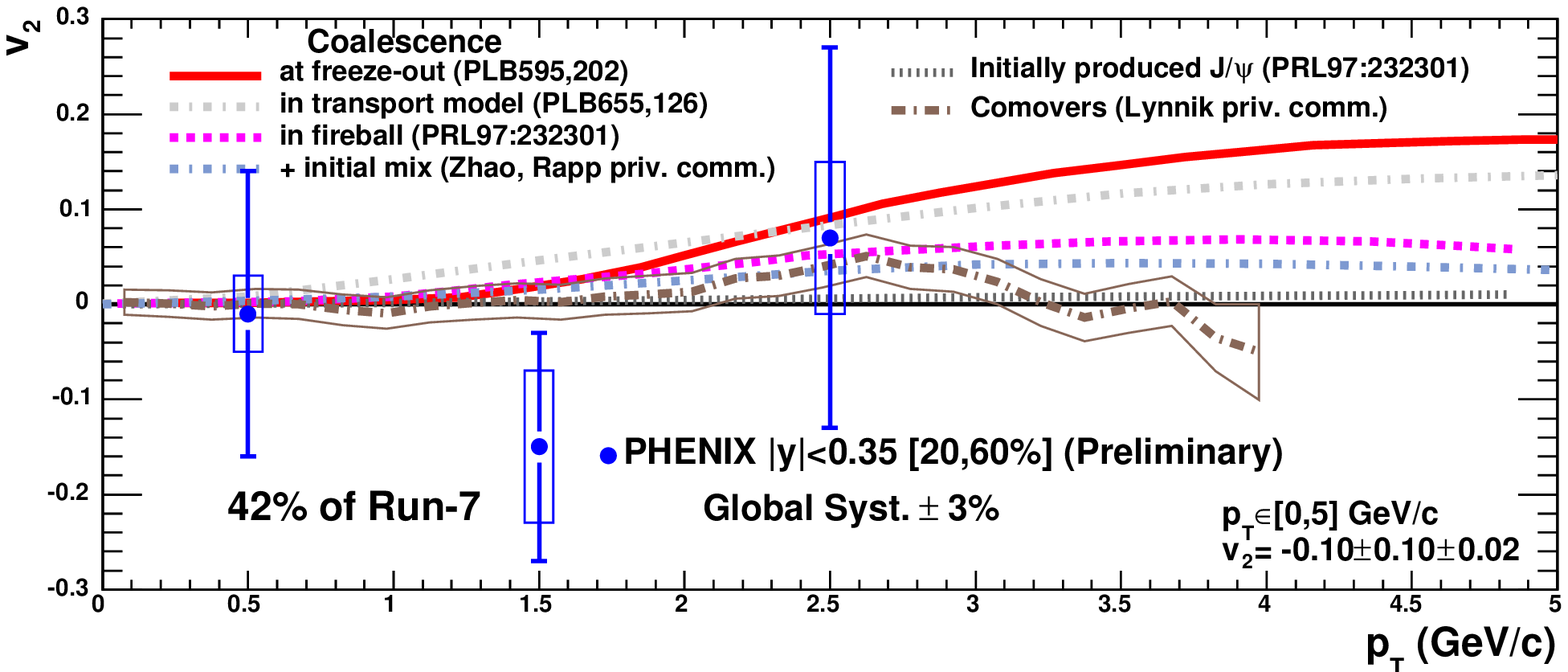} & \includegraphics[height=5cm]{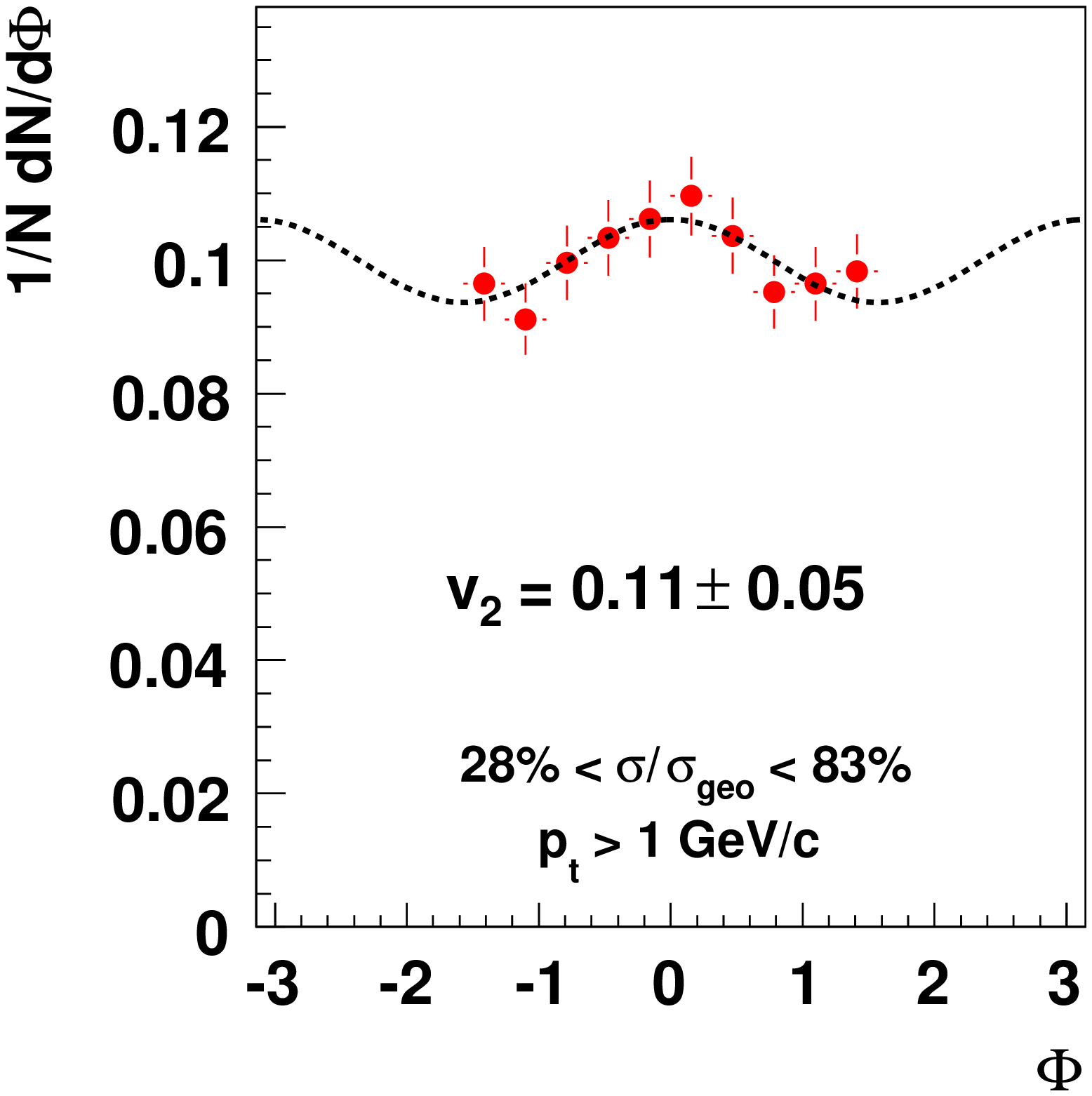}
  \end{tabular}
\end{center}
\vspace{-2ex} \caption{$J/\psi$ elliptic flow measurements. Left: $v_2$ versus $p_T$ in the PHENIX experiment, compared to various predictions (from~\cite{Silvestre:QM08}). Right: Azimuthal distribution of $J/\psi$ versus the reaction plane, exhibiting clear anisotropy in the NA60 experiment (from~\cite{Arnaldi:QM08}).
} \label{Fig4}
\vspace{-2ex}
\end{figure}

Another way of probing the coalescence hypothesis is to measure $J/\psi$ elliptic flow.
Indeed, if charmonia are produced by coalescence of charm quarks, they will somewhat inherit their flow, which we know to be quite large from open heavy flavour measurements~\cite{Adare:2006nq}. At this conference, the PHENIX experiment reported a first tentative measurement of $J/\psi$'s $v_2$, for Au-Au centrality of 20 to 60~\%~\cite{Silvestre:QM08}. As shown on the left part of figure~\ref{Fig4}, these proof-of-principle measurements do not allow one to distinguish between models assuming various regeneration (and thus elliptic flow) amount.

Statistically more significant is the result that NA60 reported, as seen on the right part of figure~\ref{Fig4}. It exhibits $v_2 = 11 \pm 5$~\% for In-In collisions of centrality 28 to 83~\%  and $J/\psi$ of transverse momentum larger than 1~GeV/$c$. At this energy and number of collisions regime, it is very unlikely to be due to elliptic flow. The number of charm quarks is low (of the order of 0.05 per average In-In collisions) and they are unlikely to thermalize. If confirmed, this anisotropy could arise from an absorption by the surrounding anisotropic nuclear matter. In any case, it makes the interpretation of a possible $J/\psi$ azimuthal anisotropy at RHIC as a sign of coalescence less obvious.

\subsection{Charmonia excited states}

The normal and anomalous suppression patterns of $\psi'$ and $\chi_c$ higher quarkonia states could help solving the $J/\psi$ case. In case nuclear effects explain the rapidity dependence of $J/\psi$ suppression, the anomalous suppression is moderate and compatible with the amount of $J/\psi$ coming from excited states decays. If so, the excited states should mostly vanish in the QGP. 
On the opposite, if regeneration is at play, excited states should also be regenerated and present after the QGP phase.  

Preliminary ratio of $J/\psi$ coming from excited states at RHIC energy were shown for the first time at this conference~\cite{Morino:QM08}. From p-p collisions, the $\psi'$ is measured, while only an upper limit at 90~\% confidence level is obtained for the $\chi_c$:
\begin{equation}
\hspace {-2em} \psi \hspace{1ex} from \hspace{1ex} \psi' = 8.6 \pm 2.5 \hspace{1ex} \% \hspace{1em} and \hspace{1em} \psi \hspace{1ex} from \hspace{1ex} \chi_c < 42 \hspace{1ex} \% \hspace{1ex} (90 \hspace{1ex} \% \hspace{1ex} ULCL)
\end{equation}

Furthermore, one starts to estimate the total beauty cross-section, which translates into a fraction of $J/\psi$ coming from $B$ at the level of $4^{+3}_{ -2}$~\%~\cite{Morino:QM08}. Similar measurements in d-Au and Au-Au collisions should bring insights on the $J/\psi$ anomalies.

\subsection{Bottomonia}

Beyond charmonia, small samples of bottomonia are now observed at RHIC. This is of particular interest, since bottomonia might be easier to understand. First, with much less than one $b\bar{b}$ pair per central Au-Au collisions, it would be very surprising if regeneration plays any role in the beauty sector at RHIC. Second, having higher masses, bottomonia are made up of higher momentum partons and will less suffer from shadowing (and may even lie in an antishadowing region).

From the first $\Upsilon$ signal shown by the PHENIX experiment three years ago~\cite{Xie:2005ac}, a measurement in A-A collisions was very awaited. At this conference, STAR has presented a $\Upsilon$ signal extracted from Au-Au collisions~\cite{Das:QM08}. Together with a companion p-p measurement, it should soon provide a nuclear modification factor. For now, the collaboration safely quotes an upper limit of $R_{AuAu} < 1.3$ at a confidence level of 90~\%.

\subsection{Onset in the suppression pattern}

From the seminal paper of Matsui and Satz~\cite{Matsui:1986dk}, the ideal signature of deconfinement would be a sudden drop of the $J/\psi$ yield, when one reaches the dissociation temperature. If such a threshold is probably observed in the NA60 data~\cite{Arnaldi:2007zz}, it is very tempting to see it on the midrapidity PHENIX data (remember the left part of figure~\ref{Fig2}). This is premature. First, such onsets need to be looked at after cold nuclear matter effects are subtracted, and we have seen in section~\ref{SecCNM} that they are poorly known at present. Second, such a discontinuous behaviour does not appear in the forward rapidity data. Last but not least, if onset models do fit the midrapidity data, so does a smooth behaviour. However, should an onset finally be seen at RHIC, it will immediately be interpreted in terms of physical quantities such as a dissociation temperature (see e.g. \cite{Gunji:QM08} at this conference).

\subsection{Quarkonia at the LHC}

Finally, even with more information on the above mentioned observables, the precise understanding of $J/\psi$ suppression at RHIC may remain delicate, the interplay between cold nuclear matter effects, anomalous suppression and regeneration being too intricate. If so, one can wonder: what will be the fate of $J/\psi$ at the LHC? The number of $c\overline{c}$ pairs will be much higher: likely larger than 100 in a single most central collision. If regeneration is at play, then the $J/\psi$ nuclear modification factor could even raise with centrality, overcoming the normal suppression and the further shadowed initial partons (see~\cite{Andronic:QM08} at this conference for an example of calculation). If so, $J/\psi$ enhancement will be a spectacular manifestation of reconfinement (and thus of deconfinement). If not, then the $J/\psi$ yields at LHC may be as hard to interpret as the ones measured at RHIC. But LHC will also broadly open the era of $\Upsilon$ measurements. The very tightly bound ground state $\Upsilon$ could be unsuppressed and thus serve as a reference\footnote{This will only be feasible once the fraction of $\Upsilon$ coming from the higher states, including $\chi_b$, is measured or at least taken into account: it is known to be as large as 50~\% at the Tevatron~\cite{Affolder:1999wm}.}
for the excited states ($\Upsilon'$ and $\Upsilon''$). This feature should ease the disentanglement of saturation and anomalous suppression in the $\Upsilon$'s family.

\section{An anomalous conclusion}

We saw that $J/\psi$ suppression is not yet understood at RHIC, in particular its rapidity dependence. To move forward, a lot of tracks can be followed, the most promising of which certainly being, on a short time scale, the analysis of the large d-Au sample to refine our knowledge of cold nuclear effects. This being said, one should keep in view that, assuming conservative cold nuclear matter approaches, some level of anomalous suppression is indeed observed at RHIC, characterizing the matter as hot and deconfining.

\section*{References}

\bibliographystyle{myunsrt}
\bibliography{Biblio}


\end{document}